# Coupled air lasing gain and Mie scattering loss: aerosol effect in filament-induced plasma spectroscopy

JIAYUN XUE[1,2,#], ZHI ZHANG[1,3,#], YUEZHENG WANG [1,2], BINPENG SHANG[1,3], JIEWEI GUO [1,2], SHISHI TAO[1,2], NAN ZHANG [1,2], LANJUN GUO[1,3], PENGFEI QI[1,2,*], LIE LIN[1,3] AND WEIWEI LIU [1,2]

[1] *Institute of Modern Optics, Eye Institute, Nankai University, Tianjin 300350, China*
[2] *Tianjin Key Laboratory of Micro-scale Optical Information Science and Technology, Tianjin 300350, China*
[3] *Tianjin Key Laboratory of Optoelectronic Sensor and Sensing Network Technology, Tianjin 300350, China,*
[#] *These authors contributed equally to this work*
*\* qipengfei@nankai.edu.cn*

**Abstract:** Femtosecond laser filament-induced plasma spectroscopy (FIPS) demonstrates great potentials in the remote sensing for identifying atmospheric pollutant molecules. Due to the widespread aerosols in atmosphere, the remote detection based on FIPS would be affected from both the excitation and the propagation of fingerprint fluorescence, which still remain elusive. Here the physical model of filament-induced aerosol fluorescence is established to reveal the combined effect of Mie scattering and amplification spontaneous emission, which is then proved by the experimental results, the dependence of the backward fluorescence on the interaction length between filament and aerosols. These findings provide an insight into the complicated aerosol effect in the overall physical process of FIPS including propagation, excitation and emission, paving the way to its practical application in atmospheric remote sensing.

## 1. Introduction

Haze is a complex mixture with carbonaceous species, heavy metals (e.g., Pb, Zn, Al, Ca, Fe, Cr and Cu) and salt compounds (e.g., sulfates, nitrates, chlorine salts and ammonium salts) [1], widely suspending in the atmosphere in the state of aerosols with the particle size of 0.01~10 µm [2, 3]. The scattering and absorbing of aerosols seriously affect the visibility and climate [4, 5]. Additionally, detrimental aerosols in atmosphere are inevitably absorbed and deposited into human body, causing difficulty in breathing, and even causing toxicity [6]. Accurate detection of aerosols is of great significance for the prevention, evaluation of haze, and the prediction of climate.

Femtosecond laser pulses with high peak power can overcome the diffraction effect and propagate over a long distance ranging from meters to kilometers[7-9] with a transverse dimension of ~100 µm in the atmosphere [10], demonstrating a great potential in the remote sensing of aerosols. The ultrastrong laser intensity ($10^{14}$ W/cm$^2$) [11] within filaments can naturally lead to ionization and fragmentation of molecules in gas, vapors, solids, and aerosols[12], inducing fingerprint fluorescence emissions[12, 13]. Among various filament-based sensing techniques, filament-induced plasma spectroscopy (FIPS) has been regarding as a versatile spectroscopic technology [14-17]. By virtue of advantages in remote, real-time, and multi-element analysis, FIPS opens the possibility of the remote sensing for identifying atmospheric pollutant molecules. Additionally, Filament-induced amplification spontaneous emission (ASE) is widely observed in $N_2$[18], $O_2$[19], CH[20], NH, OH[21], and CN[22] compounds. Furthermore, the filament is considered as the gain media to generate air lasing[19,

23-25], facilitating the improvement of the signal-to-noise ratio in filament-based remote detection technique.

The current research on FIPS, especially air lasing, has been performed at laboratory with limited propagation distance and stable gas environments. However, the Mie scattering effect is non-negligible for the nonlinear propagation of femtosecond laser[26], fluorescent emission directivity[27], as well as ASE effect due to the existence of troublesome aerosols in atmosphere. That is, the remote detection of FIPS in aerosols would be affected from both the excitation and the propagation of fluorescence. Even so, the coupled air lasing gain and Mie scattering arising from aerosols on the FIPS still remain elusive.

Here, the complete physical model was elaborately established to describe the collaborative effect of Mie scattering and amplification spontaneous emission associated with aerosols, enabling accurately reproduce the experimental results, the dependence of interaction length between filament and aerosols on backward fluorescence. These findings provide an insight into the complicated aerosol effect in the overall physical process of FIPS including propagation, excitation and emission, paving the way to its practical application in atmospheric remote sensing.

## 2. Physical model combining nonlinear wave equation with Mie theory

Femtosecond laser filamentation is a complex nonlinear phenomenon, including diffraction, group velocity dispersion, Kerr effect, multi-photon ionization, self-steepening and absorption. It can be described by the nonlinear wave equation[28, 29]:

$$-2ik_0 \frac{\partial A_0}{\partial z} = \Delta_\perp A_0 - k_0 k_2 \frac{\partial^2 A_0}{\partial t^2} + 2\left(1 + \frac{i}{\omega}\frac{\partial}{\partial t}\right)\frac{k_0^2}{n_0} n_{nl} A_0 - ik_0 \alpha_a A_0 \qquad (1)$$

where $A_0(r, z, t)$ represents the envelope of the laser amplitude, $\Delta_\perp$ is the transverse Laplacian operator, $k_0$ is the wavenumber corresponding to $\lambda_0 = 800$ nm, $k_2 = 0.2$ fs$^2$/cm is the dispersion coefficient in air, $n_0$ and $n_{nl}$ are linear and nonlinear refractive indices respectively. When the femtosecond laser propagates in aerosols, the scattering and absorption are non-negligible. It can be quantified as the attenuation coefficient of $\alpha_a = -0.205$ m$^{-1}$ according to Lambert-Beer law. A 50 cm-long filament is simulated with the input laser energy of 4 mJ and pulse width of 140 fs. As shown in Fig. 1a, the interaction length between aerosols and filament can be flexibly adjusted by moving the aerosol layer. To investigate the effect of aerosol introduced Mie scattering and filament induced ASE on the excited fluorescence, the femtosecond laser filamentation for different interaction length with aerosols is first simulated. Fig. 1b illustrates the simulated femtosecond laser filamentation for different interaction length between the filament and aerosol. Clearly, the laser energy can be significantly attenuated by the aerosols, providing increasing the interaction length from 10 cm to 40 cm, the length of filament is obviously reduced. Eventually, the femtosecond laser intensity decays so badly that lower than the critical power for filamentation due to the increased thickness of aerosol layer.

As inspired by the previous research [18, 20-22], the excited fluorescence of different ions and molecules by femtosecond laser filament undergoes amplification spontaneous emission along the filament. Therefore, the forward fluorescence signals are amplified by the gain media and absorbed by aerosol along the propagation path. The propagation of the forward fluorescence can be expressed as[23]:

$$-2ik \frac{\partial A}{\partial z} = \Delta_\perp A + 2\frac{k^2}{n_0} n_{nl} A - ik\alpha A \qquad (2)$$

where $A$ represents the fluorescence distribution, and $k$ is the wavenumber corresponding to fluorescence emission wavelength. The propagation of fluorescence is affected by the refractive index change $n_{nl}$ caused by the filament, such as the Kerr effect and plasma defocusing. The coefficient $\alpha = \alpha_g + \alpha_a$. $\alpha_a = -0.205$ m$^{-1}$ is the absorption caused by the aerosol.

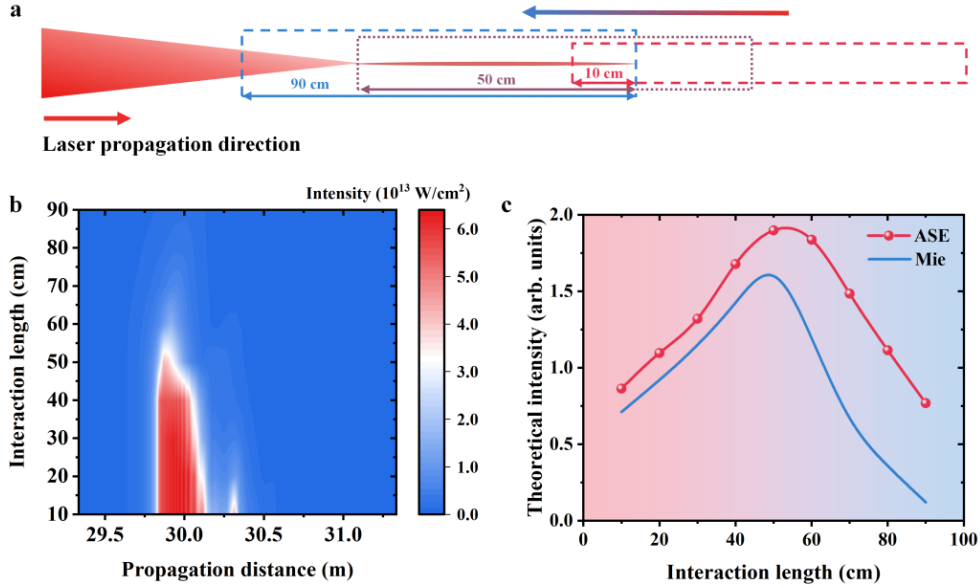

Fig. 1 (a) Schematic diagram of changing the interaction length between laser filament and aerosols. (b) Axial laser intensity inside the filament for variable interaction length between filament and aerosols. (c) Fluorescence intensity calculated by the physical model of ASE (red line) and amended by Mie scattering (blue line).

Taking the common $Na^+$ as example, $\alpha_g = N_{Na+}\sigma$ is the gain coefficient, $N_{Na+}$ is the density of $Na^+$, $\sigma$ is the transition radiation cross section of $4\times10^{-15} cm^2$ [30]. The emitted fluorescence photon counts at 589 nm can be calculated as $N_{589nm} = N_{Na+}\tau A_{21}/(e^{h\nu/kT}-1)$ [23]. The probability of spontaneous emission $A_{21}$ is $6.161\times10^7 s^{-1}$ [31]. The fluorescence lifetime $\tau$ of $Na^+$ at 589 nm is 16.2 ns [32]. The dependence of integral fluorescent intensity on interaction length can be obtained and illustrated in Fig. 1c. The fluorescence intensity climbs as the aerosol layer is forwardly moved 50 cm (the purple dotted box in Fig. 1a), due to the growth of both excited emitter and gain length. Continually moving to the position of 90 cm (the blue dash box in Fig. 1a), the aerosol layer interacts with laser before the focus, leading to the heavily attenuated fundamental femtosecond laser for filamentation, excitation and amplification.

Additionally, the backward propagating fluorescence can be affected by Mie scattering of diffused aerosols along the propagation path [27]. In details, the intensity of the Mie scattering signal depending on angular could be written as[33]:

$$I(\theta) = I_0 |S_j(\theta)|^2 / (k^2 d^2) \quad (3)$$

where $I(\theta)$ is the scattered intensity measured at distance $d$ from the scattering sphere, $I_0$ denotes the intensity of the incident light which is calculated by the fluorescence propagation equation under different interaction conditions. $k = 2\pi m_0/\lambda$ where $m_0$ and $\lambda$ represent the refractive index of the medium surrounding the scattering sphere and the wavelength of the light in a vacuum, respectively. $|S_j(\theta)|^2$ refers to the scattering coefficient in direction ($\theta$), with $j = 1$ for perpendicular polarization or $j = 2$ for parallel polarization; $\theta$ pointes to the scattering angle ($\theta = 180°$ implies backscattering). Finally, the spatial integral intensity of the backward fluorescence signals collected by the concave mirror can be calculated according to the geometric relationship of spatial collection angles. According to Eq. (3), the backward fluorescence were obtained and shown in Fig. 1c. The fluorescent intensity amended by Mie scattering is weakened in varying degrees with varying interaction length, which depends on the thickness of Mie scattering layer. The thickness of Mie scattering layer increases as the interaction length increases, leading to a stronger attenuation of the backward fluorescence. The backward fluorescent intensity still peaks at the interaction length of ~ 50 cm.

## 3. Results and discussion:

To verify the above elaborately established physical model, the dependence of interaction length between filament and aerosols on backward fluorescence were investigated in experimental. As depicted in Fig. 2a, a commercial Ti: sapphire femtosecond laser system (Legend Elite, Coherent Inc.) delivers laser pulses with the central wavelength of 800 nm, single pulse energy of 4 mJ, pulse width of 140 fs (FWHM), and repetition rate of 500 Hz. The output laser pulses are focused by a telescope system, consisting of a concave lens (L1: f = -150 mm) and concave mirror (CM: f = 2032 mm). The free-form phase plate was used to correct the aberration caused by the convergence of off-axis incident beams [34]. The filament is delivered 30 m away from the concave mirror. The filament length of 50 cm were obtained according to the filament-induced acoustic signals( Fig. 2b) [35]. $Na^+$ fluorescence is induced by the interaction of filament and NaCl aerosols. The backward fluorescence is focused by the concave mirror into the fiber bundle behind Mirror 2. A grating spectrometer (Omni-λ300, Zolix Ltd.) equipped with an Intensified CMOS camera (Istar-sCMOS, Andor Technology Ltd.) is used to analyze the backward fluorescence. The aerosols with a particle size of 1-3 μm (Fig. 1c), and the mass concentration of NaCl aerosols is 2.2 $g/m^3$ [27] are generated into the chamber (length: 100 cm, diameter: 3 cm). The interaction length of the filament and aerosol can be manipulated by moving the chamber along the propagation direction.

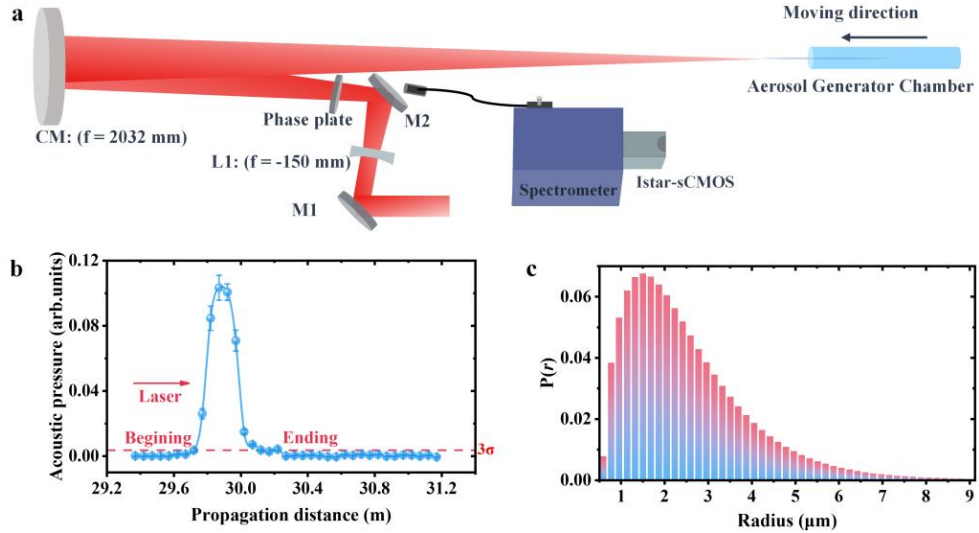

Fig. 2. (a) Experimental setup of filament-induced plasma spectroscopy system; (b) Acoustic pressure at different positions along the laser propagation direction (red line: 3 times of the standard deviation of the ambient noise 3σ); (c) Probability density function for the size distribution of the aerosol adopted in experiment.

To visualize the interaction length of the filament and aerosols, a CCD cameral is used to laterally record the filament induced $Na^+$ fluorescence emission with an interference filter (center at 589.5 nm, FWHM:<1.2 nm) to remove the scattered femtosecond laser and ambient light. Moving the aerosols chamber against the propagation direction with a step of 10 cm, the different interaction length between the filament and NaCl aerosols were obtained. Fig. 3a display the recorded lateral $Na^+$ fluorescence emitted from the filament. Clearly, the elongated longitudinal spatial distribution and enhanced fluorescence emission can be observed as increasing interaction length from 10 cm to 50 cm. When the interaction length is 50 cm, the entire filament is immersed in the aerosols. The longitudinal range and fluorescence intensity decrease providing continually moving the glass chamber to 90 cm.

According to Eq. (2), the propagation of fluorescence in aerosols is simulated and the results of the fluorescence distribution are shown in Fig. 3b. The variations in longitudinal distribution range and fluorescent intensity are similar to that of the experimental results. The fluorescence is amplified inside the filament and absorbed by the aerosols. Due to the fixed parameters of aerosol generator, the attenuation coefficient and attenuation length can be regarded as constant in all conditions. When the aerosols chamber is moved toward the starting point of the filament, the gain length is increasing. The fluorescence photons are continually generated and amplified in gain medium, resulting extended longitudinal range and enhanced fluorescence. As the interaction length is increased from 50 cm to 90 cm, the decayed laser and shortened filament leads to the decreasing gain coefficient and gain length, as well as the reduced fluorescence emission and shortened longitudinal range.

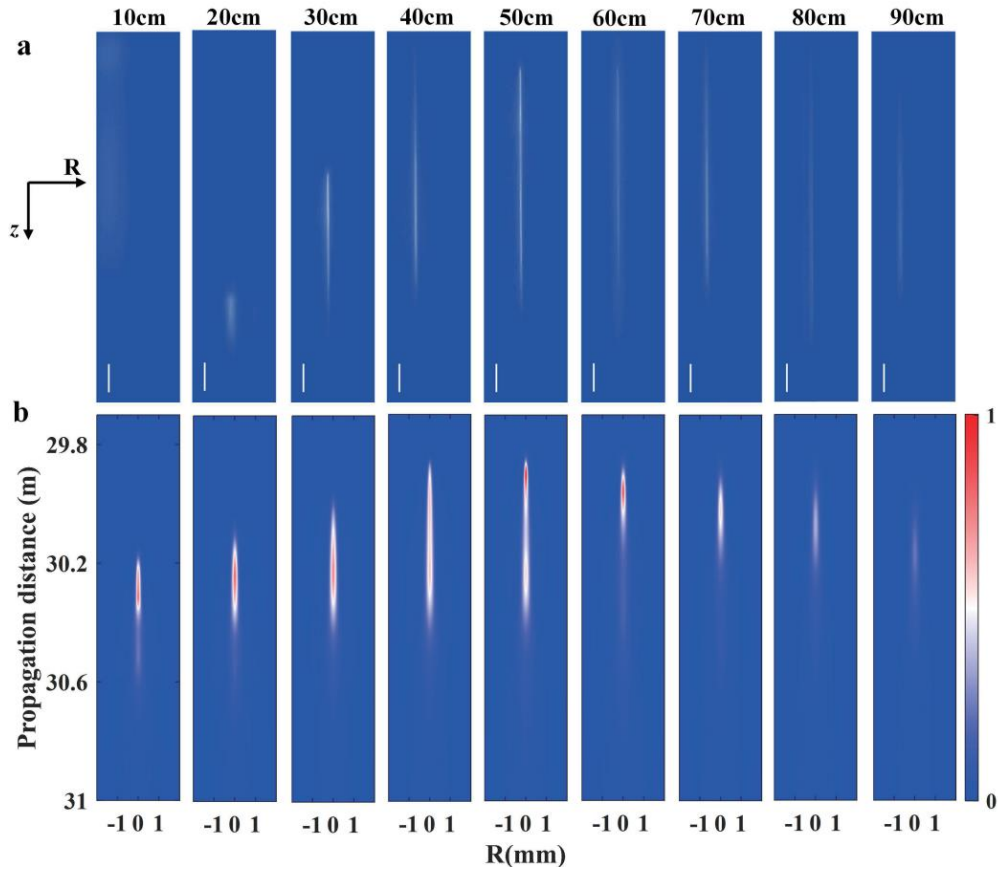

Fig.3 (a) Lateral fluorescence image from NaCl aerosols taken by CCD for different interaction length; (b) Simulated fluorescence distribution for different interaction length.

To accurately evaluate aerosol effect in filament-induced plasma spectroscopy, the backward fluorescence signals emitted from the aerosol are focused by the large concave mirror into the fiber, as shown in Fig. 4a. The integrated intensity of the fingerprint emission is summarized in Fig. 4b. The backward $Na^+$ fluorescence signal is strongest when the filament is just entirely interacting with the aerosol. The simulation results amended by Mie scattering is consistent with the trend of experimental results, as shown in Fig. 4b. The variable interaction length change the gain length and Mie scattering length. When the interaction length is less than filament length, air lasing gain plays a dominant role in the propagation of fluorescence, the intensity of fluorescent emission increases with the increasing gain length. As the interaction length is greater than filament length, the Mie scattering effect leads to a weaker excited fluorescence and stronger attenuation. As a result, the non-monotonic backward aerosol fluorescence can be attributed to the collaborative effect of Mie scattering and ASE.

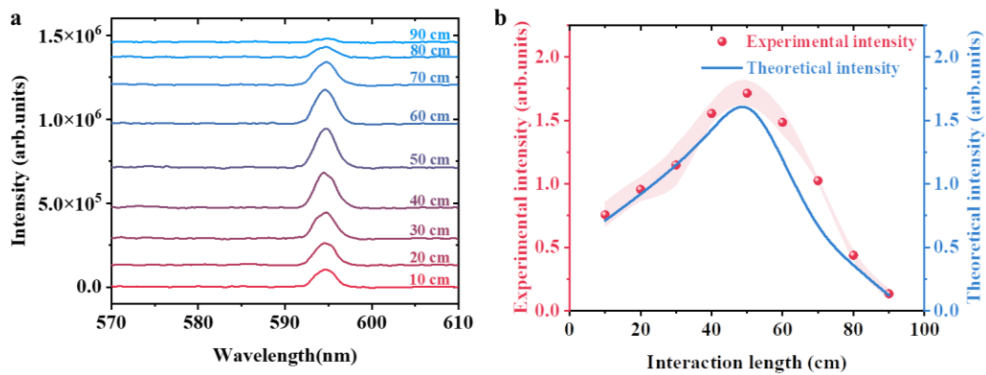

Fig.4 (a) Backward fluorescence spectra of NaCl aerosol with different interaction lengths; (b) Dependence of NaCl aerosol fluorescence intensity on the interaction length in experiment (red ball) and numerical simulation (blue line). Pink shadow: standard deviation in measurement.

## 4. Conclusion

In conclusion, the physical model of filament-induced aerosol fluorescence was built to deepen our understanding on the collaborative effect of air lasing gain and Mie scattering by aerosols. Experiment was conducted to verify the theoretical simulation results, by investigating the dependence of backward fluorescence on the interaction length between filament and aerosols. The theoretical results reproduced the experimental results. The research results lay the foundation for the application of FIPS in remote atmospheric sensing.

**Funding.** National Key Research and Development Program of China (2018YFB0504400); Fundamental Research Funds for the Central Universities (63221010, 63223052).

**Disclosures The authors declare no conflicts of interest.**